\documentclass[aps,prl,twocolumn,superscriptaddress,showpacs]{revtex4-1}
\usepackage[utf8]{inputenc} 
 
\usepackage{graphics}
\usepackage{graphicx}  
\usepackage{color}
\usepackage{float}
\usepackage{newfloat}
\DeclareFloatingEnvironment[name={FIG. S}]{suppfigure}
   
\ifx\pdftexversion\undefined
\usepackage[dvips]{hyperref}
\else
\usepackage{hyperref}
\fi
\hypersetup{
  colorlinks = true, linkcolor = blue
}
\begin{document}

\title{Single-atom-resolved probing of lattice gases in momentum space}

\author{H. Cayla}
\thanks{These authors contributed equally to this work.}
\affiliation{Laboratoire Charles Fabry, Institut d’Optique, CNRS, Université Paris-Saclay, 91127 Palaiseau cedex, France.}
\author{C. Carcy}
\thanks{These authors contributed equally to this work.}
\affiliation{Laboratoire Charles Fabry, Institut d’Optique, CNRS, Université Paris-Saclay, 91127 Palaiseau cedex, France.}
\author{Q. Bouton}
\affiliation{Laboratoire Charles Fabry, Institut d’Optique, CNRS, Université Paris-Saclay, 91127 Palaiseau cedex, France.}
\author{R. Chang}
\affiliation{Laboratoire Charles Fabry, Institut d’Optique, CNRS, Université Paris-Saclay, 91127 Palaiseau cedex, France.}
\author{G. Carleo}
\affiliation{Institute for Theoretical Physics, ETH Zurich, 8093 Zurich, Switzerland.}
\affiliation{Center for Computational Quantum Physics, The Flatiron Institute, 162 5th Avenue, New York, NY 10010, USA.}
\author{M. Mancini}
\affiliation{Laboratoire Charles Fabry, Institut d’Optique, CNRS, Université Paris-Saclay, 91127 Palaiseau cedex, France.}
\author{D. Cl\'ement}
\email[Corresponding author: ]{david.clement@institutoptique.fr}
\affiliation{Laboratoire Charles Fabry, Institut d’Optique, CNRS, Université Paris-Saclay, 91127 Palaiseau cedex, France.}

\begin{abstract}
Measuring the full distribution of individual particles is of fundamental importance to characterize many-body quantum systems through correlation functions at any order. Here we demonstrate the possibility to reconstruct the momentum-space distribution of three-dimensional interacting lattice gases atom-by-atom. This is achieved by detecting individual metastable Helium atoms in the far-field regime of expansion, when released from an optical lattice. We benchmark our technique with Quantum Monte-Carlo calculations, demonstrating the ability to resolve momentum distributions of superfluids occupying $10^5$ lattice sites. It permits a direct measure of the condensed fraction across phase transitions, as we illustrate on the superfluid-to-normal transition. Our single-atom-resolved approach opens a new route to investigate interacting lattice gases through momentum correlations.
\end{abstract}
% insert suggested PACS numbers in braces on next line
%\pacs{37.10.De, 32.80.Pj, 37.10.Gh, 05.30.Jp}
% insert suggested keywords - APS authors don't need to do this
\keywords{}

\maketitle

Ultracold atoms in optical lattices have proven to be a valuable system to investigate condensed-matter models in a tunable and controllable environment \cite{bloch2012}. In this context, the past decade has witnessed the emergence of a new generation of lattice experiments capable of measuring spatial distributions and correlations between individual particles \cite{bakr2009, sherson2010, haller2015, cheuk2015, parsons2015, edge2015, barredo2016}. These apparatus have paved the way to unprecedented investigations of both equilibrium and dynamical properties of strongly-interacting matter. Similarly, the on-site detection of individual spins is central to other experimental platforms, like trapped ions \cite{britton2012} or superconducting circuits \cite{barends2016}. Some paradigmatic manifestations of quantum many-body effects are however elusive in spatial correlations, and multi-particle correlations between other degrees of freedom play a fundamental role \cite{schweigler2017}.

In this respect, the momentum of the particles is an essential degree of freedom, whose correlation functions contain unique signatures of many-body quantum coherence. Off-diagonal long-range order and Bose-Einstein condensation manifest in the momentum density \cite{leggett2006}, and thermal and quantum fluctuations can be characterized through the population of one-particle momentum states \cite{chang2016}. While well established for weakly-interacting gases, momentum densities are notoriously difficult to measure in strongly interacting systems like liquid Helium \cite{leggett2006}. Two-particle momentum correlations are central to various microscopic mechanisms of pairing in many-body Hamiltonians, from the quantum depletion in Bose liquids to fermionic Cooper pairing \cite{leggett2006}, and they exhibit signatures of quantum phase transitions \cite{toth2008} and out-of-equilibrium dynamics \cite{chen2011}. Creating the possibility to measure the full momentum distribution of individual particles would permit a direct assessment of these many-body phenomena.

In principle, quantum gases offer the possibility to explore the momentum degree of freedom by performing time-of-flight (TOF) experiments, {\it i.e.} probing the gas after a free-fall expansion. In practice, the conditions under which the measured distributions precisely map to the in-trap momentum distributions are difficult to fulfill \cite{pollet2012}. The free-fall dynamics should not be perturbed by the presence of interactions to apply the ballistic relation between the in-trap momentum and the measured position after time-of-flight. In addition a long TOF is required to enter the far-field regime of expansion where the TOF distribution equals the Fourier transform of the in-trap $g_{1}(x,x')$ function \cite{gerbier2008}. Optical imaging is not adapted to probing extremely dilute clouds over a large volume and the TOF commonly implemented in experiments is too short to reach the far-field regime. Moreover the technical requirements to efficiently collect the photons emitted by a single atom make it difficult extending single-atom-resolved fluorescence techniques to this regime. Finally, column-integrated images smoothen, and may partially wash out, specific features of the momentum space, {\it e.g.} the sharp edges of a three-dimensional Bose condensate.

In this work, we demonstrate the ability to reconstruct, accurately and atom-by-atom, the three-dimensional momentum distribution of strongly-interacting lattice gases of metastable Helium. We compare the time-of-flight distributions measured in the experiment with the in-trap momentum distributions calculated from ab-initio Quantum Monte-Carlo (QMC). The excellent agreement shows that our approach overcomes all the issues associated with measuring momentum distributions in a TOF experiment. Accessing the momentum space allows us to introduce a precise thermometry method from the comparison with QMC calculations and to directly measure the condensed fraction $f_{c}$. Finally, we monitor $f_{c}$ across the superfluid-to-normal phase transition, illustrating the capabilities offered by our apparatus.

\begin{figure}[t!]
\includegraphics[width=\columnwidth]{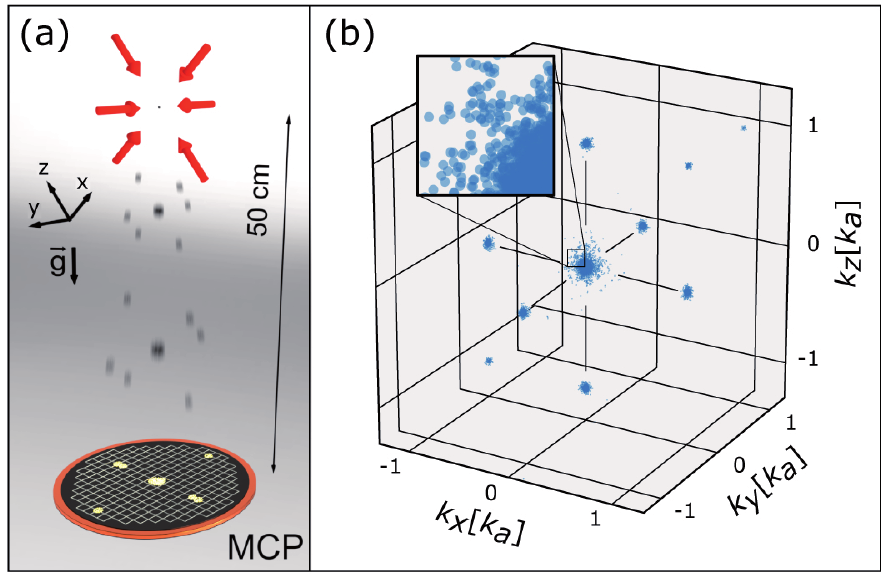}
\caption{Three-dimensional atom distributions of lattice gases in far-field. (a) Sketch of the detection method. $^4$He$^*$ atoms loaded in a 3D lattice are suddenly released and fall under gravity on the Helium detector located $\sim50$ cm beneath the trap. Atoms hit the surface of the Micro-Channel Plates (MCP) after a time-of-flight $t_{{\rm TOF}}=325\,$ms. (b) Atom-by-atom reconstruction of the 3D momentum distribution for a ratio $U/J=9.5$ of the Bose-Hubbard parameters and obtained from $\sim 500$ experimental runs (here only $t_{{\rm TOF}}=296\,$ms). Each blue dot is a single atom. The inset is a magnified view of the central region.}
\label{Fig1}
\end{figure}

Our approach exploits the properties of Helium-4 atoms brought to quantum degeneracy in a metastable state ($^4$He$^*$). On the one hand, the large internal energy of $^4$He$^*$ yields the unique possibility to detect individual atoms in three dimensions over centimeter distances \cite{vassen2012}, in contrast to optical imaging methods. On the other hand, the original combination of a long time-of-flight $t_{{\rm TOF}}=325\,$ms and the small mass of $^4$He$^*$ allows us to probe the far-field. Entering the far-field regime necessitates a TOF larger than $t_{{\rm FF}}\simeq \,m L^{2}/2\hbar$ \cite{gerbier2008}, where $\hbar$ is the reduced Planck constant, $m$ the atomic mass and $L$ the length of the trapped gas. The use of a light atomic species is thus extremely favorable to keep $t_{{\rm FF}}$ accessible, even for large system sizes $L$. For our parameters ($L\sim 50$ lattice sites), we find $t_{{\rm FF}}\simeq 50\,$ms, indeed much shorter than $t_{{\rm TOF}}$ (see \cite{SuppMat} for further details). 

\begin{figure}[t!]
\includegraphics[width=\columnwidth]{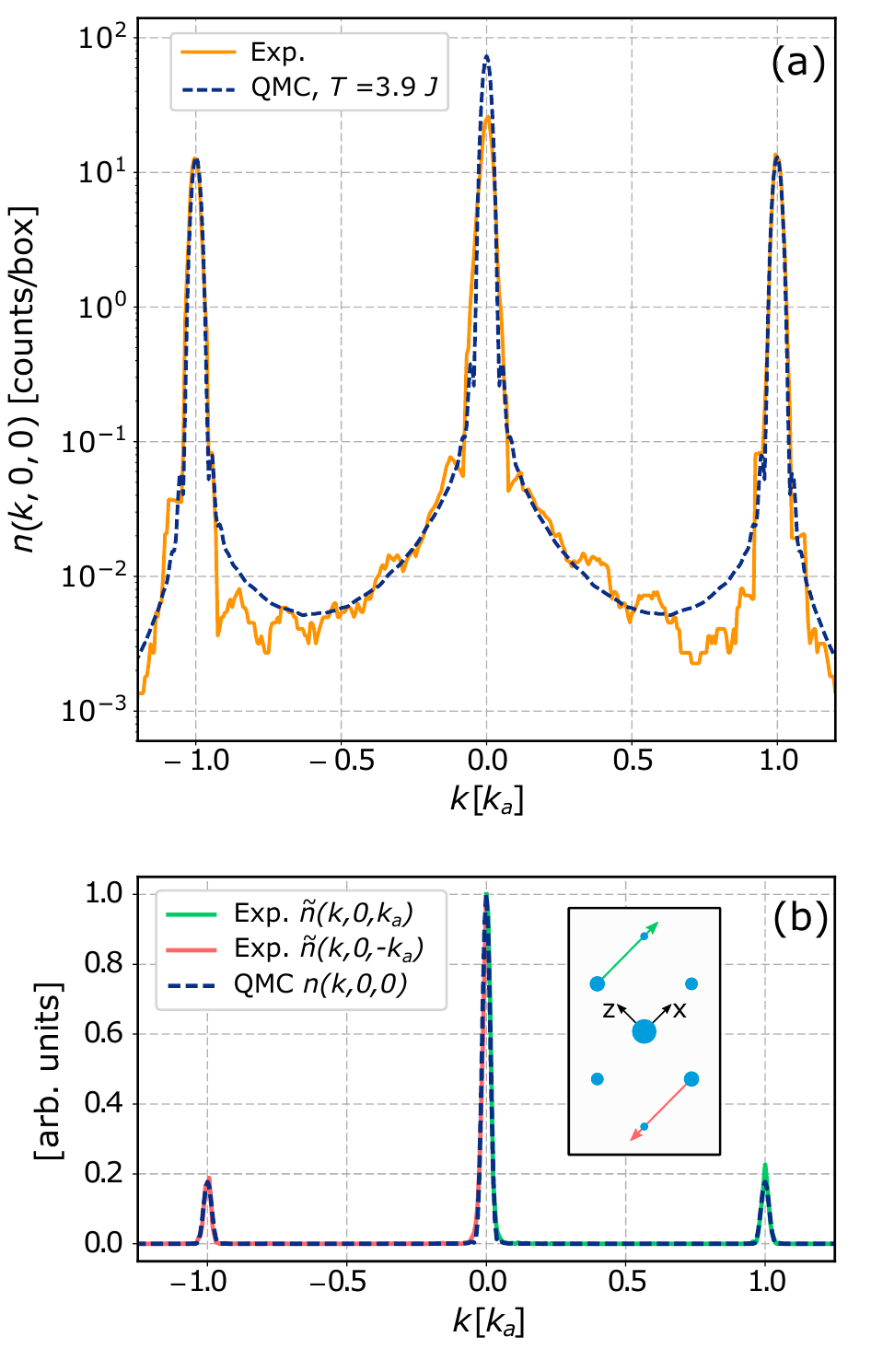}
\caption{Comparison of the measured atom distribution $n_{\rm mcp}(\vec{k})$ with ab-initio Quantum Monte-Carlo calculation of the in-trap momentum distribution $n(\vec{k})$. (a) Log-plot of 1D cuts $n_{\rm mcp}(k,0,0)$ through the measured 3D distribution (orange), and $n(k,0,0)$ through the QMC momentum distribution (dashed line) obtained for our experimental parameters and $T=3.9\,J$. The temperature $T$ is the only adjustable parameter (see text). The measured amplitude of the central peak is slightly affected by a saturation of the Helium detector (see text). (b) 1D cuts $\tilde{n}_{\rm mcp}(k,0,\pm k_{a})$ compared to the QMC profile $n(k,0,0)$. The height of the central peak of these distributions has been normalized to one. Saturation effects are not present on the first- and second-order peaks of diffraction.  \label{Fig2}}
\end{figure}

\begin{figure*}[ht!]
\includegraphics[width=2\columnwidth]{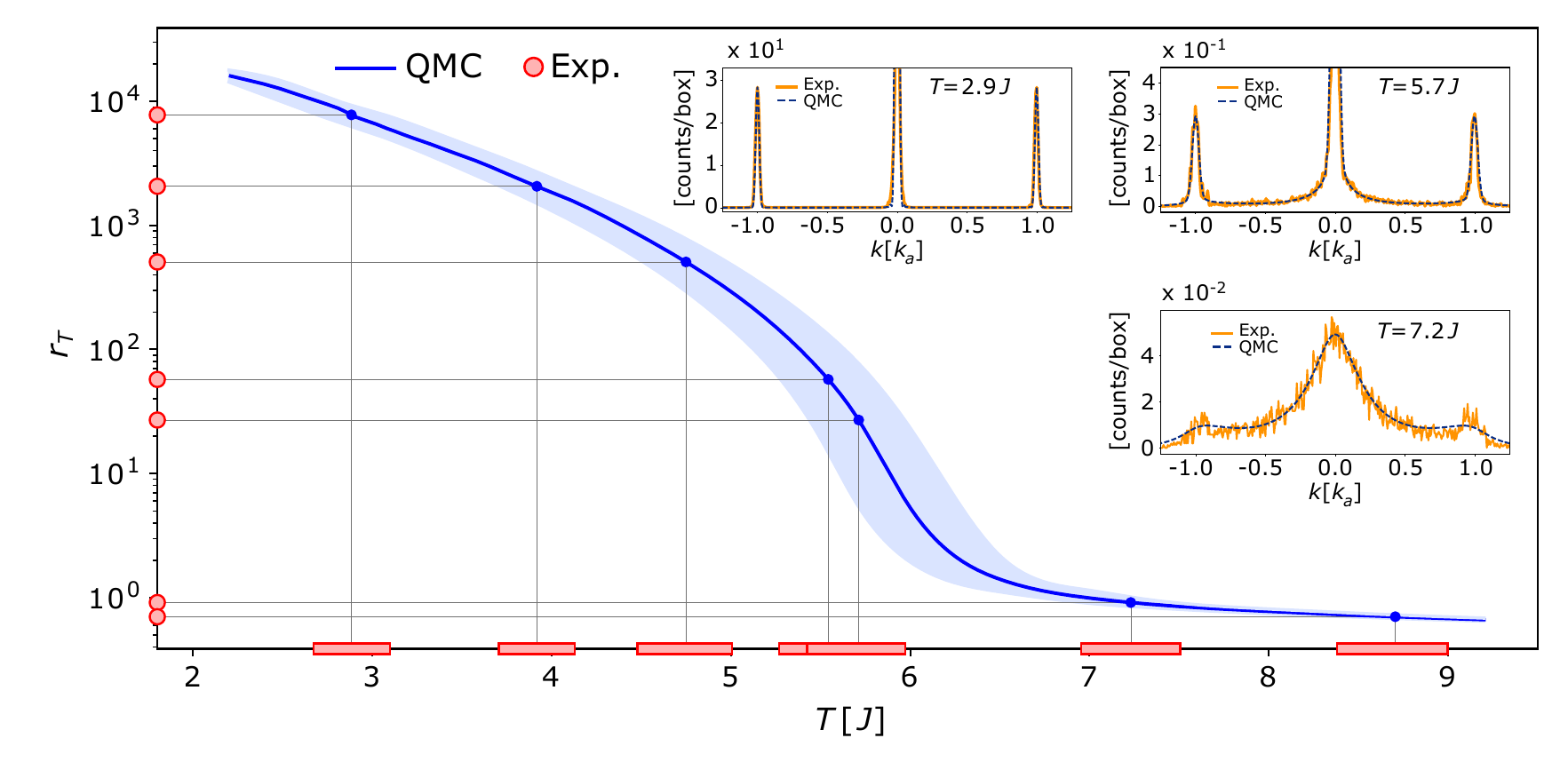}
\caption{Thermometry of lattice superfluids. Plot of the ratio $r_{{\rm T}}=n(k_{a},0,0)/n(k_{a}/2,0,0)$ as a function of the temperature $T$ as extracted from the QMC calculations (solid blue lines). The measured $r_{{\rm T}}$ (red dots) are compared with the thermometry curve to extract the temperature in the experiment (red rectangles on the temperature axis). The blue shaded area takes into account the uncertainty in the experimental atom number. The width of the rectangles depicts the corresponding error in determining the temperature. Three experimental 1D cuts $n(k,0,0)$ corresponding to $T = 2.9\,J $, $5.7\,J$ and $7.2\,J$ are also shown along with the QMC calculations.}  \label{Fig3}
\end{figure*}

We realize strongly interacting lattice superfluids by adiabatically loading a Bose-Einstein Condensate (BEC) of $ N = 40(4) \times 10^3$ $^{4}$He$^*$ atoms \cite{bouton2015} in the lowest energy band of a cubic optical lattice. The lattice spacing is $a = 775\,$nm and its amplitude $V_{\rm L} = 9.6(3) \,E_{R}$ is fixed, where $E_{R}= h^2/8m a^2$ is the recoil energy. At this amplitude $V_{\rm L}$, we have $U/J = 9.5$, where $U$ and $J$ are the Bose-Hubbard on-site interaction and tunneling energy parameters, and the expected quantum depletion is large ($\sim 15\%$), highlighting the presence of strong quantum correlations. After abruptly switching off the optical lattice, the cloud expands for $t_{{\rm TOF}}=325\,$ms and it is probed with the Helium detector (Fig.~\ref{Fig1}(a). The Helium detector yields 3D distributions of individual atoms as shown in Fig.~\ref{Fig1}(b) and has been described previously \cite{nogrette2015, chang2016}. In brief, metastable atoms hit the surface of a pair of micro-channel plates (MCPs) from which they can extract an electron by releasing their internal energy ($19.6\,$eV). The pair of MCPs multiplies the electron emitted by a single $^4$He$^*$ atom and, in combination with a delay-line anode, allows to reconstruct the 3D position $\vec{r}$ of the atom in the frame of the cloud center-of-mass \cite{chang2016}, with a detection efficiency of 25(5)\%. We associate a momentum $\hbar \vec{k}$ to the position $\vec{r}$ of an atom through the ballistic formula $\hbar \vec{k} = m \vec{r}/t_{{\rm TOF}}$, obtaining the distributions $n_{\rm mcp}(\vec{k})$ reconstructed from the MCP detector.

We now turn to the benchmarking of the measured atom distributions $n_{\rm mcp}(\vec{k})$. The theoretical framework underlying our experiment is the celebrated Bose-Hubbard Hamiltonian (BHH) \cite{trotzky2010}. Equilibrium properties of the BHH at finite temperature can be numerically computed using Quantum Monte-Carlo (QMC) approaches \cite{boninsegni2006}. This allows us to compare $n_{\rm mcp}(\vec{k})$ with ab-initio QMC calculations of the in-trap momentum distribution $n(\vec{k})$, without making any assumption about the TOF dynamics occurring in the experiment. In Fig.~\ref{Fig2}(a) we plot 1D cuts $n_{\rm mcp}(k,0,0)$ through the 3D distributions, obtained by counting the number of atoms in a box centered at $\vec{k}=(k,0,0)$ of size $(\Delta k, 3 \Delta k, 3\Delta k)$ with  $\Delta k=k_{a}/160$ where $k_{a}=2\pi/a$. The QMC calculations are performed with the experimental parameters, except for the temperature -- not measured in the experiment -- which is the only adjustable parameter \cite{SuppMat}. To take into account the efficiency of the detection process, we rescale the experimental data by matching the height of the first-order diffraction peak obtained from the QMC calculations. As shown in Fig.~\ref{Fig2}(a) we can find a temperature for which the agreement between experiment and numerics is excellent over more than 3 decades in density. Only the amplitude of the peak at $\vec{k}=\vec{0}$ is smaller than expected due to a saturation of the Helium detector at large flux of particles \cite{fraser1993}. One can avoid this effect by reducing the number of detected atoms per shot, but at the cost of averaging over larger sets of data to maintain a given signal-to-noise for $n(\vec{k})$. Interestingly, the 3D detector allows us to measure the relative amplitudes between higher-order diffraction peaks  and confirm that the saturation effect is restricted to the central peak only. To do so, we use the translation invariance of the lattice which implies that $n(k,0,\pm k_{a})=n(k,0,0) \times \tilde{w}(k_{a})$, where $\tilde{w}(k)$ is the Fourier transform of the Wannier function in the lattice. In Fig.~\ref{Fig2}(b) we plot $\tilde{n}_{\rm mcp}(k,0,\pm k_{a})=n_{\rm mcp}(k,0,\pm k_{a})/\tilde{w}(k_{a})$ and find that it compares well with the QMC calculations. The overall excellent agreement between the experiment and the QMC numerics demonstrates that we measure the in-trap momentum distribution $n(\vec{k})$ in the experiment. In addition, the outstanding resolution ($\simeq k_{a}/200$) of the Helium detector in 3D makes our apparatus capable of competing with state-of-the-art numerical methods for large systems (here $50 \times 50 \times 50$ lattice sites). 

Obtaining the in-trap momentum distribution in a TOF experiment with interacting particles may be puzzling at first. The expansion of an interacting matter-wave from a lattice of moderate (or higher) amplitude is expected to be ballistic up to small corrections \cite{gerbier2008}. This is because the zero-point energy of one site exceeds by far any other energy scale. A low lattice filling at the trap center ensures that the interaction-induced dephasing building up during the TOF can neglected as well \cite{kupferschmidt2010}. Our experiment is performed under these conditions and TOF distributions are thus expected to be similar to the in-trap momentum distributions. The excellent match we find over a large dynamical range in density (see Fig.~\ref{Fig2}) actually shows that the expansion is ballistic up to an excellent approximation and that binary collisions do not affect the far-field distributions within the experimental uncertainties \cite{ScattSphere}. In the following, we thus use the notation $n(\vec{k})$ for the measured distributions. 

In the correlated superfluid phase, two distinct contributions can be identified in the momentum distribution \cite{kato2008}, that of the condensate $n_{0}(\vec{k})$ and that of the depleted atoms $n_{{\rm NC}}(\vec{k})$ (both quantum and thermal depletion),  
\begin{equation}
n(\vec{k})=n_{0}(\vec{k}) + n_{{\rm NC}}(\vec{k}). 
\end{equation}
The condensate has long-range phase coherence and $n_{0}(\vec{k})$ has a Fourier-limited width $\propto 1/L$. On the other hand, $n_{{\rm NC}}(\vec{k})$ is a smoothly varying distribution with a typical momentum width $\propto 1/a$ given by the lattice density of state. This results from the population of all quasi-momentum states by the quantum depletion at $U/J=9.5$ and by the thermal depletion at non-vanishing temperatures. For large trapped systems, $L \gg a$, $n_{0}$ and $n_{{\rm NC}}$ thus clearly separate in the momentum space with the condensate manifesting as sharp peaks on top of the broad distribution of depleted atoms \cite{kato2008}. In the following we exploit this separation of scales to calibrate the temperature and measure the condensed fraction.

To determine the temperature, we use the comparison with QMC, inspired by the pioneering work of \cite{trotzky2010}. Here we measure a single parameter $r_{{\rm T}}=n(k_{a},0,0)/n(k_{a}/2,0,0)$ to be matched with QMC numerics. The variation of $r_{{\rm T}}$ with $T$ obtained from the QMC calculations is shown in Fig.~\ref{Fig3}. In the regime of the experiment, $r_{{\rm T}}$ varies by 4 orders of magnitude when $T$ increases only by a factor 3, yielding an unprecedented precision in the calibration of the temperature. Using a controlled heating sequence, we have varied the temperature in the experiment at a fixed $U/J=9.5$ \cite{SuppMat}. By comparing the measured $r_{{\rm T}}$ with the thermometry curve we obtain $T$ with a 5\% uncertainty,  limited by the experimental uncertainty on the atom number. In Fig.~\ref{Fig3} we also show comparisons of the measured 1D cuts $n(k,0,0)$ with QMC calculations performed at the temperature obtained from the measurement of $r_{{\rm T}}$. The excellent agreement over the entire range of temperatures validates the thermometry method. The thermometry curve in Fig.~\ref{Fig3} applies to the BHH (with our parameters) and is model-dependent. But the central idea of the method  -- probing low-energy excitations, the population of which strongly varies with the temperature -- can be generalized to a large variety of Hamiltonians. 

\begin{figure}[ht!]
\includegraphics[width=\columnwidth]{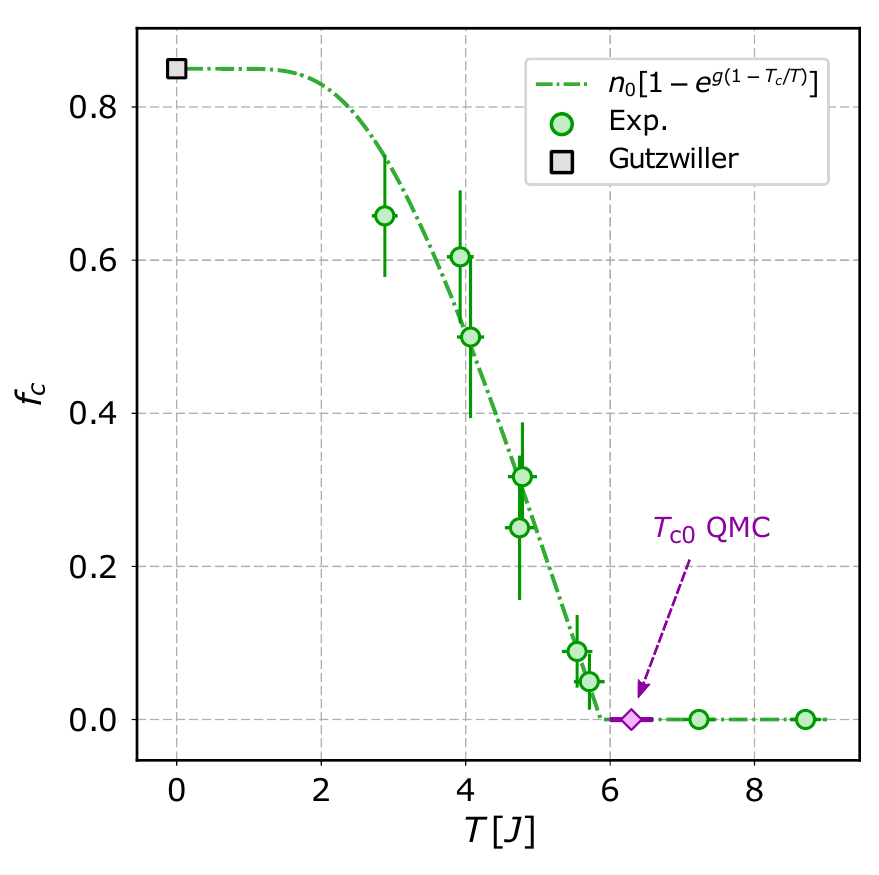}
\caption{Condensed fraction across the superfluid-to-normal phase transition. Measured condensed fraction $f_{c}$ (green dots) plotted as a function of the temperature $T$. Each point is an average over $\gtrsim 1000$ experimental realizations and the error bars correspond to one standard deviation. The mean-field prediction calculated from the Gutzwiller ansatz is shown for $T=0$ (grey square). Using an empirical fitting function (green dot-dashed line) we extract a critical temperature $T_{c}=5.9(2) \,J$ (see text). For a homogeneous lattice gas with a chemical potential matching that at the trap center, $T_{c0}=6.3(3) \,J$ (violet diamond). \label{Fig4}}
\end{figure}

The condensed fraction $f_{c}$ is a central quantity to investigate finite-temperature phase diagrams \cite{fisher1989}. But measuring $f_{c}$ in many-body systems is challenging as the strong interactions make it difficult to probe single-particle states and access $n(\vec{k})$. Techniques like neutron scattering in liquid Helium indeed fail to separate the condensate from the depleted atoms \cite{leggett2006} while Bragg spectroscopy in strongly interacting gases necessitates switching off the interactions \cite{lopes2017}. Overcoming the difficulties associated to accessing $n(\vec{k})$ \cite{pollet2012, ray2013}, our apparatus provides the first direct measurement of $f_{c}$ in a 3D lattice experiment. We identify the momentum $k_{0}$ as the momentum where the sharp profile $n_{0}$ of the condensate separates from $n_{{\rm NC}}$, finding $k_{0}\sim 0.05 k_{a}$ \cite{SuppMat}. The accurate determination of $k_{0}$ relies on probing far-field distributions in 3D since a finite time-of-flight and/or a column-integration would smoothen the density profile. We then use the full 3D distribution to count the fraction $f_{c}$ of atoms of the first Brillouin zone contained in a sphere of radius $k_{0}$ centered on $\vec{k}=\vec{0}$ and correcting for the saturation effect \cite{SuppMat}. We plot $f_{c}$ as a function of $T$ in Fig.~\ref{Fig4}. In the absence of a theoretical model, we extract the critical temperature using an empirical function \cite{ray2013}, obtaining $T_{c}=5.9(2)\, J$. Our approach allows us to observe the onset of BEC in a deep lattice with unprecedented resolution. 

An interesting question is whether a trapped gas can provide information about the phase diagram of homogeneous systems. In our experiment, BEC is expected to first appear at the trap center when $T$ is lowered from above $T_{c}$ \cite{trotzky2010}. We thus compare the measured $T_{c}$ with the critical temperature $T_{c0}$ of a homogeneous lattice gas whose chemical potential matches that at the trap center. $T_{c0}$ is obtained  from a QMC approach \cite{SuppMat} and we find $T_{c0}=6.3(3)\, J$,  a value slightly shifted ($\simeq 6 \%$) but compatible with $T_{c}$. This observation suggests that our apparatus is suited to explore trap-size scaling and criticality \cite{campostrini2009, pollet2012}. The direct measure of $f_{c}$ also opens novel perspectives to investigate many-body quantum phase transitions, {\it e.g.} many-body localization in disordered lattices \cite{aleiner2010}. 

In conclusion, we have demonstrated to measure the 3D momentum distributions of interacting lattice gases with unprecedented accuracy and single-atom resolution. We have shown to estimate the temperature within a few \% and to directly measure the condensed fraction. Thanks to single-atom resolution, the Helium detector can in principle access momentum correlations at any order. While its low detection efficiency ($\simeq$ 25\%) could be a limitation, momentum correlations up to 6th-order were previously measured in ideal Bose gases \cite{dall2013}. In the future, it should thus be possible to probe multiple-particle correlations across quantum phase transitions, to allow for an experimental characterization of the many-body ground-state wave-function undergoing the transition \cite{amico2008} or to monitor the dynamics of momentum correlations after a quench \cite{chen2011}.  

%%%%%%%%%%%%%%%%%%%%%%%%%%%%%%%%%%%%%%%%%%%%%%%%%%%%%%%%%
\vspace{0.5cm}
\begin{acknowledgments}
We acknowledge fruitful discussions with A. Aspect, T. Bourdel, M. Cheneau, M. Holzmann, P. N. Ma, L. Sanchez-Palencia, M. Troyer and the members of the Atom Optics group at Institut d'Optique. We acknowledge financial support from the LabEx PALM (Grant number ANR-10-LABX-0039), the International Balzan Prize Foundation (2013 Prize for Quantum Information Processing and Communication awarded to A. Aspect), the Institut Francilien de Recherche sur les Atomes Froids, the ERC Advanced Grant "SIMCOFE" (FP7/2007-2013 Grant Agreement No. 290464), and the Swiss National Science Foundation through NCCR QSIT. Numerical simulations were performed on the CSCS Mönch cluster. D.C. is a member of the Institut Universitaire de France.
\end{acknowledgments}

%%%%%%%%%%%%%%%%%%%%%%

\newpage

\cleardoublepage

\section{Supplemental material}
\setcounter{figure}{0}
\renewcommand{\thefigure}{S\arabic{figure}}
\renewcommand{\bibnumfmt}[1]{[S#1]}

\subsection{Experimental Sequence} 

Almost pure BECs of about $4 \times 10^4$ atoms in the state $2^3S_1$, $m_J=1$ are initially prepared in a crossed optical trap (ODT) as described in \cite{bouton2015}. We adiabatically load the BEC in the lowest energy band of an optical lattice made of three pairs of counter-propagating laser beams derived from a high-power laser source at telecom wavelength 1550 nm (Keopsys CEFL-KILO Series 15W). The lattice beams are turned on with a $100\,$ms-long exponential ramp with characteristic time $\tau=20\,$ms. Meanwhile, the crossed dipole trap is switched off with an exponential ramp of duration 80 ms and $\tau=20\,$ms. The final trapping frequencies are given by the gaussian shape of the lattice beams and equal to $\vec{\omega}/2 \pi \approx (308, 295, 298)\,$Hz. In order to test the adiabaticity of the process we revert the loading procedure and check that the BEC possess no discernible thermal fraction. 

The atoms are left in the optical lattice for a variable holding time (depending on the presence or not of the heating procedure) after which all laser beams are switched off and the cloud expands under gravity for a free fall of 50 cm. During the first $100\, \mu s$ of expansion we rapidly transfer a variable fraction of atoms to the non-magnetic state $2^3S_1$, $m_J = 0$ with a resonant RF pulse ($\Delta E = h \times 6.8$ MHz). We then apply a magnetic gradient to push the atoms remaining in the $m_J\,=\,\pm 1$ states away from the detector. The RF pulse power, combined with its time duration allows us to control the flux of atoms ($m_J = 0$) reaching the detector. We typically operate between 1 and $50\,\%$ RF transfer efficiency, the latter being chosen for the datasets at higher temperature ($T > 7\,J$) for which the atomic samples are more dilute. For the low temperature gases where the momentum density spans several orders of magnitude, we reconstruct the 1D profiles from merging low and high flux data as introduced in \cite{chang2016}.

\subsection{Calibration of the single-atom Helium detector}

The Helium detector is made of a pair of Micro-Channel Plates (Burle Industries, channel diameter $25\,\mu$m, center-to-center spacing $32\,\mu$m, angle of the channel $8^\circ$) mounted onto a crossed delay-line anode (Roentdek DLD80). The electronic pulses at the four ends of the delay line are amplified and converted into NIM pulses with dedicated constant fraction discriminators (Roentdek company). The digital coding of the arrival times is obtained thanks to a home-made time-to-digital converter described in \cite{nogrette2015}.

To calibrate the efficiency of the Helium detector we have used dilute thermal clouds released from an harmonic optical dipole trap. The atom number in the thermal clouds is first measured by absorption imaging with a 10$\%$ accuracy. We then monitor the detected atom number on the Helium detector as a function of the radio-frequency amplitude and duration. We have found a detection efficiency of the Helium detector equal to $25(5)\%$. 

The accuracy with which the position of an atom is reconstructed with the Helium detector has different origins in the plane of the MCPs and orthogonally to the MCP surface. The in-plane accuracy is limited by electronic noise and jitters and we have calibrated it by measuring a quantity $\sigma_{D}$ that we have shown to provide a quantitative answer \cite{nogrette2015}. We have measured an in-plane resolution of $60 \, \mu$m corresponding to $0.002 k_{a}$. The accuracy along the axis of gravity, orthogonal to the MCP surface, is limited by the geometry of the micron-size channels of the MCP. A measure of this vertical resolution is difficult but an upper limit can be obtained assuming an equal detection efficiency over the entire aperture of the channels. It yields an upper value equal to $0.008 k_{a}$.

When the atomic flux on the Helium detector is too high, the micro-channels do not have enough time to reload properly, leading to lower detection efficiencies. This saturation of the MCP is a complex and long-standing phenomena \cite{fraser1993}. In our experiment, we observe a saturation for local fluxes of a few MHz/cm$^2$. For the parameters we use, only the central diffraction peak is affected by this saturation, which leads to a suppression of the atom density around $k\simeq 0 $. The side-peaks are instead not saturated, thanks to the envelope of the Wannier function which decreases the local flux of particles by a factor of 0.17 for a lattice depth $s = 9.6\, E_{R}$. This is confirmed in Fig.~\ref{Fig2}b where $\tilde{n}_{\rm mcp}(k,0,\pm k_{a})$ is shown to be in excellent agreement with the QMC calculations.

\subsection{Reaching the far-field regime of expansion}

To test the validity of the far-field approximation when using the Helium detector, we have investigated the time-of-flight dynamics of lattice BECs. The latter yields an interference pattern that can be pictured by considering the lengths $a$ and $L\gg a$ associated to the trapped BEC, giving rise to momentum components $k_{a}=2 \pi/a$ and $2 \pi/L$ respectively. The momentum $k_{a}$ is associated with the distance between diffraction peaks in the interference pattern and $2 \pi/L$ to the width of these peaks. As larger momenta develop faster during TOF, the diffraction peaks quickly separate. On the contrary, it takes longer for the width to establish. The far-field regime is reached when all momentum scales are fully developed, which happens after a TOF $t_{{\rm FF}}$ that can be expressed in terms of the smallest momentum component, $t_{{\rm FF}}= \,m L^{2}/2\hbar$ \cite{gerbier2008}. To monitor the relative dynamics of the momentum scales $2 \pi/L$ and $k_{a}$, one can measure the ratio of the RMS peak size $\sigma$ to the peak-to-peak separation $d$ as a function of $t_{{\rm TOF}}$. In far-field where the interference pattern is fully developed, the ratio $\sigma/d$ should be constant and proportional to $a/L$. 

\begin{figure}[ht!]
\includegraphics[width=\columnwidth]{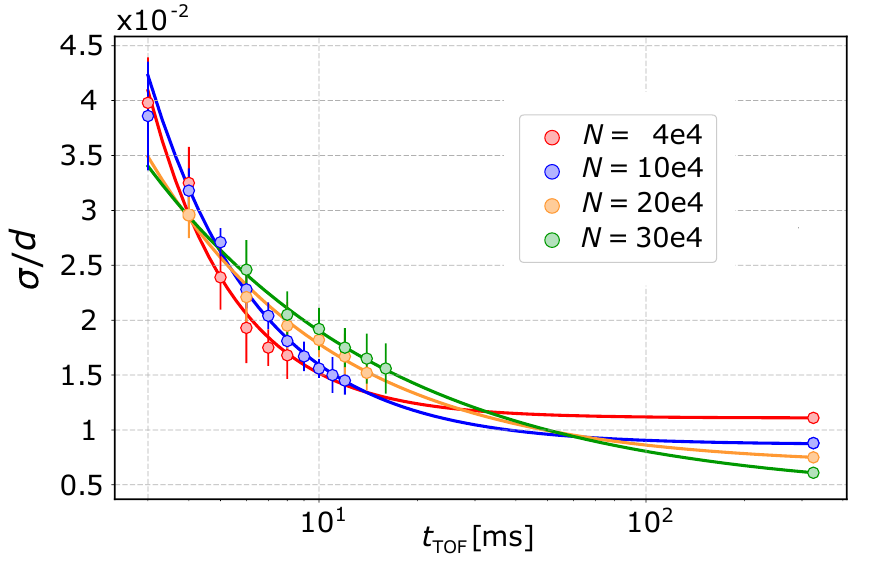}
\caption{Reaching the far-field regime during the TOF expansion of lattice superfluids. Time evolution of the width $\sigma$ of the diffracted peaks normalized to the peak distance $d$. Data sets with a varying total atom number $N$ correspond lattice superfluids with different in-trap size $L$. Solid lines are a guide to the eye. \label{SupFig1}}
\end{figure}

In the experiment, we let the the cloud expand for a varying TOF and it is probed using two methods. At short TOF (from 2 to $20\,$ms) we take 2D absorption images. After a long TOF of $325\,$ms, we use the Helium detector. As shown in Fig.~\ref{SupFig1}, we observe that $\sigma/d$ indeed saturates at long TOF, after a fast decrease at short TOF. In addition, we have further verified the variation with the system size $L$ (at fixed $a$) by increasing the atom number $N$ in the condensate to increase $L$. As expected, we observe that, the smaller the system size $L$, the faster the initial TOF dynamics and the larger the final value $\sigma/d$ (see Fig.~\ref{SupFig1}). For $N=4 \times 10^4$ atoms we find that $t_{{\rm FF}}\simeq 50\,$ms is much shorter than $t_{{\rm TOF}}=325\,$ms, indicating that the distributions measured by the Helium detector is in far-field. 

\subsection{Heating sequence}

To vary the temperature in the lattice, we have used a controlled heating sequence. After loading the BEC in the lattice, a series of non-adiabatic pulses of $0.5\,$ms duration each is performed with one of the lattice beams. The number of pulses (from 5 to 10) and their amplitude $V_{\text{heat}}$ are varied to obtain different final temperatures in the lattice while keeping the number of atoms constant. After the pulse sequence, the cloud is left to thermalize for a time interval of $40\,$ms ($ \sim 20\,h/J$). By exploiting a band mapping technique we have verified that no discernible fraction of atoms is transferred to higher lattice bands.  
  
\subsection{QMC calculations}

\renewcommand{\citenumfont}[1]{S#1}

Exact thermodynamic properties of interacting lattice bosons can be studied using Quantum Monte Carlo (QMC) methods. In this work we have used the worm algorithm QMC, following the scheme of Pollet et al. \cite{pollet2007}. 

The main idea of this approach is to consider a stochastic representation of the partition function $Z=\mathrm{Tr}[e^{-\beta H}]$, with $\beta=1/k_{b}T$
is the inverse temperature, and $H$ is the Bose-Hubbard Hamiltonian. The starting point is to decompose the Hamiltonian into two terms:
$H_{0}=\frac{U}{2}\sum_{i}n_{i}(n_{i}-1)+V_{\mathrm{trap}}$ and $H_{1}=-J\sum_{\langle i,j\rangle}b_{i}^{\dagger}b_{j}+\mathrm{h.c.}$,
and consider the Dyson series for the partition function: 
\begin{eqnarray*}
Z & =\sum_{n=0}^{\infty}(-1)^{n}\underbrace{\int d\tau_{1}\dots d\tau_{n}}_{0<\tau_{1}\leq\dots\leq\tau_{n}<\beta}\mathrm{Tr}e^{-\beta H_{0}} & H_{1}(\tau_{n})\dots H_{1}(\tau_{1}),
\end{eqnarray*}
where $H_{1}(\tau)=e^{\tau H_{0}}H_{1}e^{-\tau H_{0}}$ is the time-evolved
$H_{1}$ in the interaction representation. Each of the terms in the Dyson series are then sampled stochastically, using an extended configuration space including open world lines (the so-called ``worm''). 

This approach also allows to measure efficiently the equal-time, two-particle Green function $g_{lm}=\langle b_{l}^{\dagger}b_{m}\rangle$, for pairs of sites $l$ and $m$. The momentum distribution for the trapped system is then obtained as:
\begin{eqnarray*}
n(\vec{k}) & = & |w(\vec{k})|^{2}S(\vec{k}),
\end{eqnarray*}
where $w(\vec{k})$ is the Fourier transform of the Wannier function associated to the optical lattice, and 
\[
S(\vec{k})=\sum_{lm}g_{lm}e^{i\vec{k}\cdot(\vec{R}_{l}-\vec{R}_{m})},
\]
where $\vec{R}_{l}$ denotes the coordinate of the $l-$th lattice site. In our simulations, we have fixed the total chemical potential in the system in order to match the estimated experimental value for the total number of particles in the system. 

For the homogeneous lattice gas (i.e. in the absence of a trapping potential, and considering a cubic box with periodic boundary conditions) we have determined the transition temperature $T_{c0}$ for the superfluid transition through the winding number estimator of the superfluid density \cite{ceperley1995}. In this case the chemical potential has been fixed to match the chemical potential estimated at the center of the trap for the experimental configuration. Our calculations were performed using an updated version of the DWA code in the ALPS library \cite{alps2007, alps2011}. 
   
\subsection{Extracting the condensed fraction $f_{c}$} 

As explained in the main text, the inequality $L\gg a$ allows us to distinguishing the condensate contribution $n_{0}(\vec{k})$ from that of the depletion $n_{{\rm NC}}(\vec{k})$. The condensate $n_{0}(\vec{k})$ consists in sharp peaks of size $\propto 1/L$ and measuring the condensed fraction $f_{c}$ thus amounts to measuring the atom number $N_{\rm peak}$ in the condensate peaks with respect to the total atom number $N$, $f_c=N_{\rm peak}/N$. Here we make the assumption that $N_{\rm peak} \simeq N_{0}$ with $N_{0}$ the number of condensed atoms and $N = N_{0} + N_{{\rm NC}}$. This also relies on the inequality $a/L \ll1$ which ensures that the fraction of depleted atoms contained in the sharp peaks is extremely low, of order $(a/L)^3$.

\begin{figure}[ht!]
\includegraphics[width=\columnwidth]{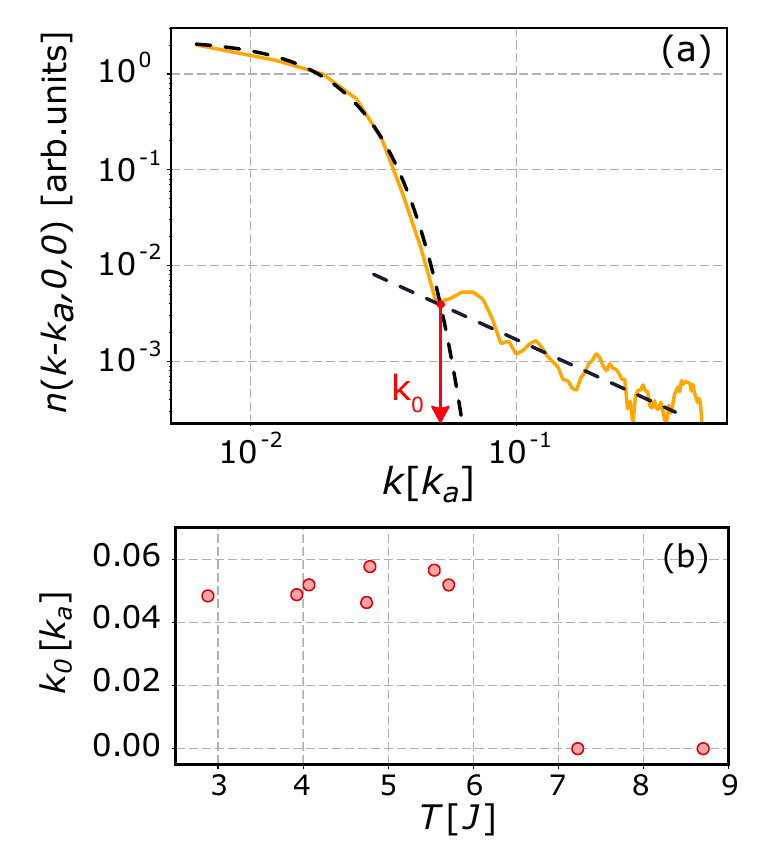}
\caption{Extracting $k_{0}$ from experimental profiles. (a) Finding $k_{0}$ as the momentum at which a Gaussian fit of the diffraction peak intersects with a power-law fit of the depletion. (b) $k_{0}$ as a function of $T$.   \label{SupFig2}}
\end{figure}

To identify the momentum $k_{0}$ which delimits the two contributions to $n(\vec{k})$ we focus on the first diffracted peak that is not affected by saturation effects of the detector. We fit 1D profiles with a Gaussian centered at $k=k_a$ on top of a power-law. We defined $k_{0}$ as the momentum where these two fitted distributions intersect, as illustrated in Fig.~\ref{SupFig2}(a). The large dynamical range in density provided by the Helium detector is of primary importance to accurately fit the tails associated to the depletion of the condensate and identify $k_{0}$ accurately. A similar approach would result difficult from column-integrated images as obtained when probing quantum gases with absorption imaging. In Fig.~\ref{SupFig2}(b), we plot $k_{0}$ as a function of the temperature $T$. For $T>T_{c}$, one can not identify a double structure with an abrupt change of slope (see experimental profiles at $T=7.2\,J$ in Fig.~\ref{Fig3}) and we set $k_{0}=0$.
  
Since the momentum distribution is translation invariant (up to an envelop factor given by the Fourier transform of the Wannier function), we extract $f_{c}$ by considering a single Brillouin zone only. In the 3D distributions, we count the atom number in the central peak $N_{\rm peak}^0$ and the atom number in the first Brillouin zone $N_{{\rm BZ}}$. To take into account the saturation of the detector around $k\simeq 0$, the measured atom number in the central peak is multiplied by a scaling factor $\alpha$ obtained from the amplitude of the first-order diffracted peak, $\alpha = n(k_{a},0,0)/[\tilde{w}(k_{a})\cdot n(0,0,0)]$. We then calculate $f_{c}=\alpha \cdot N_{\rm peak}^0/N_{{\rm BZ.}}$.

\end{document}